\documentstyle{article}

\begin{document}

\title{Generalization of the Berezinskii-Kosterlitz-Thouless theory
to higher vortex densities}
\author{Carsten Timm \\
Universit\"at Hamburg, I. Institut f\"ur Theoretische Physik, \\
Jungiusstra\ss{}e 9, D-20355 Hamburg, Germany}
\date{February 23, 1996}

\maketitle

\begin{abstract}
The Berezinskii-Kosterlitz-Thouless theory for superfluid films
is generalized in a straightforward way that
(a)~corrects for overlapping vortex-an\-ti\-vor\-tex pairs at high pair
density and (b)~utilizes a dielectric approximation for the polarization
of the vortex system and a local field correction.
Generalized Kosterlitz equations are
derived, containing higher order terms, which are compared with earlier
predictions. These terms cause
the total pair density to remain finite for temperatures
above the transition so that it is not necessary to
introduce an {\it ad hoc\/} cut-off, as opposed to the original
Berezinskii-Kosterlitz-Thouless theory.
The low-temperature bound pair phase is destabilized for small
vortex core energy. The behaviour of the stiffness constant
and of the correlation length
close to the transition is not affected by the higher order terms.
A first-order transition as suggested by other authors
is not found for any values of the parameters.
The pair density is calculated for temperatures below and above
the transition. Possible experiments and the applicability of the
extended approach are discussed. The approach is found to be applicable
even in a significant temperature range above the transition.
\end{abstract}

\section{Introduction}
\label{sec:intro}

Two-dimensional and quasi-two-dimensional systems with a two-component
order parameter and short-range interactions typically fall
into the universality class of the two-dimensional $XY$ model.
The behaviour of such systems, e.g., superconducting films, superfluid
films and thin planar magnets, close to their phase transition is usually
well described by the Berezinskii-Kosterlitz-Thouless (BKT)
theory \cite{Bere,KT}.
The discovery of the high-temperature superconductors, which consist of
weakly coupled superconducting layers, has led to renewed
interest in this theory \cite{HTKT}.
In BKT theory the phase transition is supposed to take
place through the unbinding of spontaneously created
vortex-antivortex pairs. For temperatures
up to the transition temperature $T_c$ the predictions of BKT theory
typically agree quite well with experimental results. There are several
essentially equivalent mathematical formulations of the BKT
theory \cite{KT,Kost74,Jose,Young,Minn}.
Here, we follow Halperin's
presentation \cite{Halp}, which is founded upon the heuristic
approach of Kosterlitz and Thouless \cite{KT}.

The BKT theory treats the interaction between different
vortex-antivortex pairs through a suitable renormalization of the intra-pair
interaction.
This approximation is best justified if pairs are small and far apart.
The so-called Kosterlitz equations, which are derived in BKT theory,
relate the density of vortex-antivortex pairs of
given size and the screening of the vortex-antivortex interaction at given
separation to one another. Nelson and Kosterlitz \cite{NK} and Halperin
\cite{Halp} predict the general form of higher order terms of these
equations, but do not derive the coefficients.
Amit {\it et al}.\ \cite{Amit}. utilize field-theoretical methods to derive
a universal relation between the coefficients of two higher order terms.
One would like to know the higher order terms, including the
coefficients, to see if the typical BKT behaviour,
which is found in experiment, survives in their presence.

Without these terms the BKT theory suffers from the following
weakness: If taken literally it predicts a diverging vortex density
for $T>T_c$. This result cannot be correct. The creation of one
vortex-antivortex pair costs at least twice the core energy. The
density of pairs should thus be governed by a Boltzmann factor containing
this energy. In the BKT theory, a cut-off length $\xi_+$ is introduced
{\it ad hoc\/} to obtain finite results. The length $\xi_+$ is meant to
represent the average separation between free vortices. In a system of zero
overall vorticity, all vortices and antivortices can be grouped into
pairs. Thus $\xi_+$ is not very well-defined.
The concept of ``free vortices'' does not naturally fit into the
theory. It is unsatisfactory that such a cut-off is necessary at all.
It seems that an important contribution to the density has been neglected.

Different objections are put forward by Gabay and Kapitulnik \cite{GK}.
These authors argue that one should use the Clausius-Mossotti
formula \cite{Kitt} for the dielectric constant to take
local field effects into account.
Several authors \cite{MiWa,Zhang,Ding} predict the breakdown of the BKT
theory for small vortex core energy and the appearance of a first-order
transition.
These papers are inspired by Monte Carlo simulations \cite{Caillol,Tei},
which indicate the presence of such a transition for low vortex core energies.
Minnhagen and Wallin \cite{MiWa} employ a linear screening theory to
obtain generalized Kosterlitz equations. It seems that local field
effects are not included in their approach (see below)
and that the restriction that vortices may only be created in neutral pairs
is not taken into account.
Furthermore, the authors
use different sets of self-consistency equations below and above the critical
temperature, whereas one would welcome an approach which predicts
a phase transition without having to assume its presence beforehand.
In Refs.~\cite{Zhang} and \cite{Ding} phase diagrams are derived
that do not match the BKT result for high core energies,
$E_c>q^2$, where the BKT theory should be applicable. Here, $q^2$ is the
coupling constant appearing in the unscreened vortex-antivortex
interaction $V(r)=q^2\ln(r/r_0)$.
Experimental evidence for a first-order transition does not seem to exist as
of yet.

In the following sections, a generalized BKT theory is proposed,
which is based on straightforward physical ideas and
takes the points mentioned above into consideration.
A geometric correction, which avoids the diverging density, is
employed. For the derivation of the screening of the intra-pair interaction,
we start from an exact expression of linear response theory and
then take the polarization of the vortex system and local field effects
into account.
We end up with higher order terms in both
Kosterlitz equations. These terms are compared with the predictions mentioned
above and the generalized equations are discussed. It is shown that
these equations describe the vortex system even in a significant
temperature range above the transition.

\section{Geometric correction at high pair density}
\label{sec:geo}

Let $\nu(r)\,dr$ be the areal density of vortex-antivortex pairs
of sizes between $r$ and $r+dr$.
Then the total density of pairs is
\begin{equation}
n = \int_{r_0}^\infty dr\,\nu(r) ,
\label{def.n}
\end{equation}
where $r_0$ is the minimum pair size, which is of the order of the
Ginzburg-Landau coherence length.

The BKT ansatz for the pair density is
\begin{equation}
\nu(r)\,dr = \frac{N_0^2}{r_0^4} \,2\pi r\,dr\,\exp(-2\beta E_c)
  \exp[-\beta V_{\mbox{\scriptsize eff}}(r)] ,
\label{BKT.nu}
\end{equation}
where $N_0$ is the number of possibilities to place a vortex in an area
$r_0^2$, $\beta=1/T$ is the inverse temperature
(we set $k_B=1$ throughout the paper),
$E_c$ is the core energy of a vortex and $V_{\mbox{\scriptsize eff}}$
is the effective intra-pair interaction.
The expression for the pair density contains a geometric factor
$2\pi r\,dr$, which represents the area of the circular ring
in which the separation vector resides, and a Boltzmann factor
$\exp(-\beta[2E_c+V_{\mbox{\scriptsize eff}}(r)])$, which contains
the energy of the pair.

We now understand why the total density diverges for $T>T_c$.
In this temperature regime we expect the effective interaction to be
screened, i.e.\
$\lim_{r\to\infty} V_{\mbox{\scriptsize eff}}(r) = \mbox{const}$.
Thus $\nu(r)\,dr \propto r\,dr$
for large $r$. Equation~(\ref{def.n}) then implies that the total density
$n$ diverges with the system size squared.
As noted above, this result cannot be correct since the energy of a
pair is at least $2E_c>0$. Hence, an important effect must
have been neglected in
Eq.~(\ref{BKT.nu}). As explained in the following, the
error lies in the enumeration of pairs.

The idea of BKT theory is to group the vortices and antivortices into
neutral pairs and to describe the interaction between different pairs
through the renormalization of the intra-pair interaction.
However, we must not group the vortices and antivortices arbitrarily.
For the approximation to be best justified,
we must form pairs as small as possible \cite{Halp}.
We now imagine that we fill the system with pairs, starting from the
smallest ones (of size $r_0$). The probability that we add a pair of
size $r$
contains the geometric term and the Boltzmann factor present in
Eq.~(\ref{BKT.nu}). Furthermore, we must make sure that the
enumeration remains consistent. If we placed the vortex of the new pair
closer than a distance $r'$ to the antivortex of another pair, where
$r'$ is the size of the latter pair, the new vortex and the old antivortex
should be considered a pair. In that case, we would have miscounted
the pairs. Our ansatz is to introduce an additional factor
in the density, which denotes the probability that
the pairs are placed properly.
This probability is determined by the distribution
of the smaller pairs already present.

A new pair of size $r$ be tentatively
placed anywhere in the system.
The probability that the new vortex is not placed closer than $r'$ to
the antivortex of any pair of given size $r'$ is
$1-\nu(r')\,dr'\,\pi r'^2$, since $\nu(r')\,dr'$ is the density of such pairs.
The probability that the new vortex is not placed closer than $r'$ to
the antivortex of {\it any\/} pair of size $r'<r$ is then given by
\begin{eqnarray}
\prod_{r'<r} \left[1-\nu(r')\,dr'\,\pi r'^2\right]
  & = & \prod_{r'<r} \exp\left[-\nu(r')\,dr'\,\pi r'^2\right] \nonumber\\
& = & \exp\left[-\pi \int_{r_0}^r \!dr'\,r'^2\nu(r')\right] .
\end{eqnarray}
The probability that the placement of the new pair does not violate the
enumeration rule is just the square of the last expression since a
corresponding condition must be satisfied for the antivortex. Thus we obtain
\begin{eqnarray}
\nu(r)\,dr & = & \frac{N_0^2}{r_0^4} \,2\pi r\,dr\,\exp(-2\beta E_c)
  \exp[-\beta V_{\mbox{\scriptsize eff}}(r)] \nonumber\\
& & \times \exp\left[-2\pi \int_{r_0}^r \!dr'\,r'^2\nu(r')\right] 
\label{gen.nu}
\end{eqnarray}
instead of Eq.~(\ref{BKT.nu}).

This integral equation for $\nu(r)$ can be solved by transformation into
a differential equation for $Y(r) = \int_{r_0}^r dr'\,r'^2\nu(r')$.
The solution is
\begin{equation}
\nu(r) = \frac{2\pi r n^\ast \exp\left[-\beta
  V_{\mbox{\scriptsize eff}}(r)\right]}
  {1+4\pi^2n^\ast \int_{r_0}^r\!dr'\,r'^3 \exp\left[-\beta
  V_{\mbox{\scriptsize eff}}(r) \right]} ,
\end{equation}
where $n^\ast = N_0^2/r_0^4\,\exp(-2\beta E_c)$.

However, Eq.~(\ref{gen.nu}) is more suitable for the derivation
of the corresponding generalized Kosterlitz equation. Employing the fugacity
\begin{equation}
y(r) = \sqrt{\frac{r^3\nu(r)}{2\pi}}
\label{def.y}
\end{equation}
and the logarithmic length scale $l = \ln(r/r_0)$ as
defined in the BKT theory, we obtain, by taking the logarithm of
Eq.~(\ref{gen.nu}) and differentiating with respect to $l$,
\begin{equation}
\frac{dy}{dl} = \left(2-\frac{\beta}{2}
  \frac{dV_{\mbox{\scriptsize eff}}}{dl}\right) y - 2\pi^2 y^3 .
\end{equation}
Defining the dielectric constant as the ratio of bare and effective forces,
\begin{equation}
\epsilon = \frac{dV/dr}{dV_{\mbox{\scriptsize eff}}/dr}
  = \frac{dV/dl}{dV_{\mbox{\scriptsize eff}}/dl} ,
\end{equation}
assuming a logarithmic form of the potential, $V=q^2\ln(r/r_0) = q^2 l$,
and defining the stiffness constant of BKT theory by
\begin{equation}
K(l)=\frac{\beta q^2}{2\pi\epsilon(l)} ,
\label{def.K}
\end{equation}
we eventually find
\begin{equation}
y' = (2-\pi K)y - 2\pi^2 y^3 .
\label{gen.Ko.y}
\end{equation}
This equation contains the additional term $-2\pi^2y^3$ as compared with the
original Kosterlitz equation. The corresponding initial condition,
\begin{equation}
y_0 \equiv y(l=0) = N_0 \exp(-\beta E_c) ,
\label{Ko.ini1}
\end{equation}
is the same as in BKT theory since the geometric correction
results in a factor of unity in Eq.~(\ref{gen.nu}) for the smallest pairs.

The additional term in Eq.~(\ref{gen.Ko.y}) is proportional to $y^3$ as
predicted by Hal\-pe\-rin \cite{Halp}. It is also of the same
form and has the same sign as the term found by
Amit {\it et al}.\ \cite{Amit}.
Note that the notation of Ref.~\cite{Amit} differs from the
usual one of BKT theory \cite{Halp}.
The implications of this term are discussed in Sec.~\ref{sec:disc}.
Here, we only note that since $y\ge0$ by definition
and $y'<0$ for sufficiently large $y$,
the fugacity $y$ is bounded for any $K$, whereas $y$
diverges for $T>T_c$ in the original BKT theory. Equation~(\ref{def.y})
then implies that the density
$\nu$ falls off at least as fast as $r^{-3}$ for large $r$ so that the total
density $n$ also remains finite.

\section{Polarization and local field correction}
\label{sec:pol}

The question arises of whether higher order terms have also to be included
in the second Kosterlitz equation in order to consistently describe
higher vortex densities. To address this question we first rederive the
equation for the stiffness coefficient $K$ starting from linear response
theory and then discuss the necessary changes due to finite vortex densities.
In the present section we use the language of the
Coulomb gas model, i.e., we view the vortices as charged hard-core particles
interacting via the two-dimensional Coulomb potential $V = q^2 \ln(r/r_0)$.

In linear response theory the dielectric function
$\epsilon({\bf k}) = V({\bf k})/V_{\mbox{\scriptsize eff}}({\bf k})$
is derived,
\begin{equation}
\frac{1}{\epsilon({\bf k})} = 1- \frac{2\pi\beta}{k^2}\,g({\bf k}) ,
\label{pol.eps}
\end{equation}
where $g({\bf k})$ is the Fourier transform of the charge
(vorticity) density correlation function,
\begin{equation}
g({\bf r}) = \langle \rho(0)\rho({\bf r}) \rangle .
\end{equation}
The macroscopic dielectric constant is defined as
\begin{equation}
\frac{1}{\epsilon_\infty} \equiv \lim_{k\to 0} \frac{1}{\epsilon({\bf k})} .
\end{equation}
Assuming that $1/\epsilon_\infty$ is finite, we see from Eq.~(\ref{pol.eps})
that $\lim_{k\to0} g({\bf k})$ and $\lim_{k\to0} dg({\bf k})/dk$ have to
vanish so that \cite{Minn85,Alas}
\begin{eqnarray}
\frac{1}{\epsilon_\infty} & = & 1-2\pi\beta\,\frac{1}{2}\,\left.\frac{d^2g}{dk^2}
  \right|_{k\to0} \nonumber \\
& = & 1-\pi\beta \int d^2r\, g(r)\left.\frac{d^2}{dk^2} e^{-i{\bf k}\cdot
  {\bf r}}\right|_{k\to0} \nonumber \\
& = & 1+\pi\beta \int_0^\infty\!\! dr\,r \int_0^{2\pi}\!\! d\phi\, g(r)
  r^2\cos^2\phi \nonumber \\
& = & 1+\pi^2\beta \int_0^\infty\!\! dr\,r^3 g(r) .
\label{pol.einf}
\end{eqnarray}
Here, we always have $1/\epsilon_\infty<1$.

One now defines a scale-dependent dielectric constant
\cite{Minn85,Alas},
\begin{equation}
\frac{1}{\epsilon(r)} = 1 + \pi^2\beta \int_{r_0}^r \!\! dr'\,
  r'^3 g(r') ,
\label{pol.er}
\end{equation}
which includes the contribution of pairs of sizes $r'\le r$ only.
Since we employ a hard-disk model we have introduced
a lower cut-off $r_0$ in the integral. Note that $\epsilon(r)$ is not
the Fourier transform of $\epsilon({\bf k})$ as given by Eq.~(\ref{pol.eps}).

Since the correlation function $g$ is not known exactly 
we need an approximation for this quantity. As discussed above, we can always
decompose the vortex system into neutral pairs. We expect the correlation
function to consist of three terms: (a)~A $\delta$-function term denoting the
autocorrelation. This term is irrelevant because of the lower cut-off $r_0$.
(b)~A term from the partner of the test vortex, which should be proportional
to the pair size distribution $\nu(r)$. (c)~A term from the polarization
of the other pairs. Here, we make the usual
assumption that only smaller pairs are
polarized in the field of a large one \cite{Young}.
Neglecting the autocorrelation term, we write
\begin{equation}
g(r) = g_{\mbox{\scriptsize pair}}(r) + g_{\mbox{\scriptsize pol}}(r) .
\end{equation}
The relation between $g_{\mbox{\scriptsize pair}}$ and
the pair density $\nu$ is given by
\begin{equation}
g_{\mbox{\scriptsize pair}}(r) = - \frac{q^2\nu(r)}{\pi r} ,
\end{equation}
taking into account that $\nu$ has already been integrated over the angle.
To determine $g_{\mbox{\scriptsize pol}}$ we employ a dielectric approximation. We assume
that the smaller pairs form a dielectric gas characterized by the
dielectric constant $\epsilon(r)$, which is to be obtained
self-consistently. The appearance of $\epsilon(r)$ instead of $\epsilon_\infty$
is a result of the approximation that only pairs smaller
than $r$ contribute to the screening.

If two test charges, $q$ at $0$ and $-q$ at ${\bf r'}$,
are brought into a neutral medium with dielectric
constant $\epsilon(r')$, the {\it total\/} charge density, including the
polarization charges, is given by
\begin{equation}
\rho_t({\bf r}) = \frac{q}{\epsilon(r')}\,\delta({\bf r})
  - \frac{q}{\epsilon(r')}\,\delta({\bf r-r'}) .
\end{equation}
In the present approach the effect of screening is to reduce the charges of
a pair of size $r$ to the effective values $\pm q/\epsilon(r)$. Thus,
the effect of screening on the correlation function can be expressed as
\begin{equation}
g(r) = g_{\mbox{\scriptsize pair}}(r) + g_{\mbox{\scriptsize pol}}(r)
  = - \frac{q^2\nu(r)}{\pi r\,\epsilon^2(r)} .
\label{pol.appr.g}
\end{equation}
{}From Eq.~(\ref{pol.er}) we get
\begin{equation}
\frac{1}{\epsilon(r)} = 1 - \pi\beta q^2 \int_{r_0}^r \!\! dr'\,
  r'^2\,\frac{\nu(r')}{\epsilon^2(r')} .
\label{pol.appr.e}
\end{equation}
Employing the definitions (\ref{def.y}) and (\ref{def.K}), we obtain
\begin{equation}
\frac{2\pi K}{\beta q^2} = 1 - \frac{8\pi^4}{\beta q^2} \int_0^l \!\! dl'\,
  y^2(l') K^2(l')
\end{equation}
and, after differentiating with respect to $l$,
\begin{equation}
K' = -4\pi^3 y^2 K^2 .
\end{equation}
This equation is identical to the one obtained in BKT theory.

In the above considerations the smaller vortex-an\-ti\-vor\-tex pairs are
treated as a {\it continuous\/} polarizable medium. In particular,
local field effects are neglected in
the correlation function given by Eq.~(\ref{pol.appr.g}).
Consequently, the dielectric constant of Eq.~(\ref{pol.appr.e})
does not include these effects, either.
Here, we employ a Clausius-Mossotti type formula to obtain a
better approximation for $\epsilon$.
In two dimensions the relation between the dielectric constant neglecting
local field effects, $\epsilon_a$, and the full dielectric constant
$\epsilon$ is given by
\begin{equation}
\epsilon = \frac{1+\epsilon_a}{3-\epsilon_a} .
\end{equation}
Taking $\epsilon_a$ from Eq.~(\ref{pol.er}) we thus make the ansatz
\begin{eqnarray}
\frac{1}{\epsilon(r)} & = & \frac{1+\frac{3\pi}{2}\,\beta\int_{r_0}^r \! dr'\,
  r'^3 g(r')}{1+\frac{\pi}{2}\,\beta\int_{r_0}^r \! dr'\,r'^3 g(r')}
  \nonumber \\
& = & \frac{1-\frac{3\pi}{2}\,\beta q^2 \int_{r_0}^r \! dr'\, r'^2
  \nu(r')/\epsilon^2(r')}{1-\frac{\pi}{2}\,\beta q^2 \int_{r_0}^r \!
  dr'\, r'^2 \nu(r')/\epsilon^2(r')} ,
\end{eqnarray}
where $g(r)$ is not the (unknown) exact correlation function, but the one
without local field corrections, Eq.~(\ref{pol.appr.g}).
Inserting the definitions (\ref{def.y}) and (\ref{def.K})
and expanding for small $y$, we obtain
\begin{eqnarray}
K & = & \frac{\beta q^2}{2\pi} - 4\pi^3 \int_0^l \!\! dl'\, y^2(l') K^2(l')
  \nonumber \\
& & {}- \frac{16\pi^7}{\beta q^2} \left[\int_0^l \!\! dl'\, y^2(l') K^2(l')
  \right]^2 .
\label{pol.K3}
\end{eqnarray}
Differentiation yields
\begin{equation}
K' = -4\pi^3 y^2 K^2 \left[1+\frac{8\pi^4}{\beta q^2}
  \int_0^l dl' y^2(l') K^2(l') \right] .
\label{pol.dK}
\end{equation}
This expression is similar to the predicted one \cite{NK,Halp}:
The second term on the right hand side is of fourth order in $y$.
However, it is not
simply proportional to $y^4$.  Note that this result is qualitatively
different from the one of Amit {\it et al}.\ \cite{Amit}. These authors
effectively perform an expansion for a different small parameter,
namely $K-2/\pi$ instead of $y$, which makes direct comparison difficult.

To second order in $y$, the term in brackets in Eq.~(\ref{pol.dK}) is
given by $2-2\pi K/\beta q^2$, as can be seen from Eq.~(\ref{pol.K3}).
Thus we find, to fourth order in $y$,
\begin{equation}
K' = -4\pi^3 y^2 K^2 \left(2-\frac{2\pi K}{\beta q^2}\right) .
\label{pol.dK2}
\end{equation}
For pairs of minimum size $r_0$ there is no screening ($\epsilon=1$) and,
therefore, the initial condition reads
\begin{equation}
K_0 \equiv K(l=0) = \frac{\beta q^2}{2\pi} .
\label{Ko.ini2}
\end{equation}
Equation (\ref{pol.dK2}) can thus be rewritten as
\begin{equation}
K' = -4\pi^3 y^2 K^2 \left(2-\frac{K}{K_0}\right) .
\end{equation}

\section{Discussion of the generalized equations}
\label{sec:disc}

The generalized Kosterlitz equations now have the form
\begin{eqnarray}
K' & = & -4\pi^3 y^2 K^2 \left(2-\frac{K}{K_0}\right) ,
\label{Ko1a} \\
y' & = & (2-\pi K)y - 2\pi^2 y^3 ,
\label{Ko2}
\end{eqnarray}
together with the initial conditions given by Eqs.~(\ref{Ko.ini1}) and
(\ref{Ko.ini2}). We proceed to discuss these equations.
Note that the equation for $K'$ contains the initial value $K_0=q^2/2\pi T$
as an additional parameter so that the derivatives at a point $(K,y)$
are not exclusively determined by $K$ and $y$, as opposed to the original
BKT theory.

We are interested in the behaviour of Eqs.~(\ref{Ko1a}) and (\ref{Ko2})
on large length scales,
$l\to\infty$. The equations have an
attractive fixed point at $K=0$, $y=1/\pi$ and an attractive line of
fixed points at $K\ge2/\pi$, $y=0$. The first fixed point corresponds to
a screened interaction at large distances ($K\to 0$); it is the
high-temperature fixed point. Contrary to BKT theory, however, 
where this fixed point is at $K=0,\: y=\infty$, the
fugacity remains finite. We note here that our approach, which
essentially treats all vortices as bound in pairs and thus has no
free vortices, predicts the correct, metallic form of screening
for the high-temperature phase. Of course one has to check whether the
decomposition of the vortex system into pairs is justified for $T>T_c$.
This is done below. The line of fixed points corresponds to a
vanishing density of large pairs ($y\to 0$) together with a logarithmic
interaction at large distances ($K$ remaining finite). This is the
bound-pair phase.

In BKT theory analytical expressions for the trajectories, i.e.\ for the
curves $(K(l),y(l))$, can be found.
The same is not true for Eqs.~(\ref{Ko1a}) and (\ref{Ko2}).
However, these equations are not difficult to integrate numerically.
Since the equations contain the additional parameter $K_0$, every
initial point now defines its own trajectory.
In Fig.~\ref{fig:tra} trajectories for $E_c/q^2=0.5$
and several temperatures (thin solid lines) as well as for $E_c/q^2=0.2$ and
several temperatures (dashed lines) are plotted. The dotted lines denote
curves of initial values for $E_c/q^2=0.2$ and $E_c/q^2=0.5$,
respectively, and varying
$T/q^2$. The trajectories are shown in $K$-$y$ space instead of the
more usual $x$-$y$ space, where $x=-1+2/\pi K$, to emphasize the
high-temperature behaviour. The trajectories in $x$-$y$ representation
are similar to the ones of standard BKT theory.

An important question is which initial values belong to which phase,
or put mathematically, which points belong to the basin of attraction
of the high-tem\-pe\-ra\-ture fixed point and of the line of low-temperature
fixed points,
respectively. As in the original BKT theory, the two basins of attraction are
separated by the initial values which flow to $K=2/\pi,\: y=0$ in the
limit $l\to\infty$. These points form the so-called separatrix, which can
be obtained numerically and is shown in Fig.~\ref{fig:tra} as the thick
solid line. The separatrix is found to lie below the BKT result, i.e.,
the high-temperature phase is stabilized. This result is not surprising
since local field effects tend to increase $\epsilon$, thereby weakening
the renormalized vortex interaction.

Although the trajectories themselves cannot be obtained analytically,
limiting forms of the separatrix and the leading temperature dependence
of $K_\infty=K(l\to\infty)$ for $K_0\approx 2/\pi$ and $y\ll 1$ can be
deduced from the recursion relations (\ref{Ko1a}) and (\ref{Ko2}).
First, we consider the renormalized stiffness coefficient $K_\infty$.
For $T>T_c$ the flow is to the high-temperature fixed point and we have
$K_\infty=0$. At the transition we recover the universal value
$K_\infty=2/\pi$ of BKT theory. To find the behaviour for $T<T_c$, we
consider Eqs.~(\ref{Ko1a}) and (\ref{Ko2}) to leading order only,
\begin{eqnarray}
K' & = & -16\pi y^2 ,
\label{KL1} \\
(y^2)' & = & -2\pi y^2\left(K-\frac{2}{\pi}\right) .
\label{KL2}
\end{eqnarray}
These equations lead to
\begin{equation}
K'' = -2\pi K' \left(K-\frac{2}{\pi}\right)
\end{equation}
with the initial conditions $K(0)=\beta q^2/2\pi$ and $K'(0) = -16\pi y_0^2$.
The solution is
\begin{equation}
K = \frac{2}{\pi} + \sqrt{\frac{C}{\pi}}\mbox{coth}\!\left(
  \sqrt{\pi C}l + \mbox{arcoth}\!\left[\sqrt{\frac{\pi}{C}}\left(
  K_0-\frac{2}{\pi}\right) \right]\right)
\end{equation}
with $C = \pi(K_0-2/\pi)^2 - 16\pi y_0^2$. In the limit $l\to\infty$ we thus
obtain
\begin{equation}
K_\infty = \frac{2}{\pi} + \sqrt{\left(K_0-\frac{2}{\pi}\right)^{\!2}
  - 16y_0^2} .
\label{Kinf4}
\end{equation}
We know from the analysis of the fixed points that $K_\infty=2/\pi$ at
$T=T_c$. Therefore, to the present order the radicand in Eq.~(\ref{Kinf4})
vanishes at the transition. To the same order we thus find
\begin{equation}
y_0 \cong \frac{1}{4}\!\left(K_0-\frac{2}{\pi}\right)
\end{equation}
as the limiting form of the separatrix, which is the same result
as given by BKT theory.

{}From the initial conditions (\ref{Ko.ini1}) and (\ref{Ko.ini2}) we
see that the leading temperature dependence of the radicand in
Eq.~(\ref{Kinf4}) is linear so that
\begin{equation}
K_\infty \cong \frac{2}{\pi} + 4\,\sqrt{B\,\frac{T_c-T}{q^2}}
\label{Kinf5}
\end{equation}
for small $T_c-T\ge 0$. Here, $B$ is some function of the core energy
independent of temperature. Equation (\ref{Kinf5}) also has the same
form as in BKT theory \cite{Halp,Minn}. The typical square-root cusp
and the universal jump are thus left unchanged by the higher order terms.

The same is true for the typical length scale $\xi_+$, which is interpreted
as the correlation length of the superconducting order parameter. This
length is also used as the aforementioned cut-off in BKT theory. Here,
it does not have such a significance and we only discuss it for
completeness. We define this quantity as the length scale on which the
trajectories start to be drawn towards the high-temperature fixed point,
i.e., we define $K(r=\xi_+) = 2/\pi$. The length scale $\xi+$ thus
diverges for $T\to T_c$ from above. For $T<T_c$ the high-temperature
fixed point never comes into play and we have $\xi_+=\infty$. Under
the same assumptions as above and starting again from Eqs.~(\ref{KL1})
and (\ref{KL2}) we find
\begin{equation}
\ln \frac{\xi_+}{r_0} \cong \frac{1}{\pi\sqrt{16 y_0^2 - (K_0-2/\pi)^2}} .
\end{equation}
Since the radicand is again linear in $T$ to leading order, we reobtain
the BKT result
\begin{equation}
\xi_+ \cong r_0 \exp\!\left(\frac{b\,q}{\sqrt{T-T_c}}\right)
\end{equation}
for small $T-T_c > 0$ with $b$ independent of temperature.

The stability of the well-established predictions for $K_\infty$ and $\xi_+$
in the presence of higher order terms lends additional support to
the general concept of BKT. As far as these quantities are concerned,
however, both the original theory and the extension presented here are
in agreement with experiment. Thus, experiments such as measurements
of the current-voltage characteristics of superconducting films, which
effectively measure $K$, cannot distinguish between the two approaches.

Such a distinction may be possible, however,
using the total pair density $n$. In Sec.~\ref{sec:geo} we already
mentioned that $y$ is bounded in the present case and that, therefore,
$n$ must remain finite. We now discuss this point in more detail.

{}From Eqs.~(\ref{def.n}) and (\ref{def.y}) the total density
in natural units is given by
\begin{equation}
nr_0^2 = 2\pi \int_0^\infty \!dl\,e^{-2l} y^2(l) .
\label{n.def2}
\end{equation}
This integral cannot be evaluated analytically since $y^2(l)$ is unknown.
One can, however, obtain a rigorous statement
concerning the continuity of $n$. Substitution of $u=e^{-l}$ for $l$ yields
\begin{equation}
nr_0^2 = 2\pi \int_0^1\!du\,u\,y^2(u) .
\label{n.def3}
\end{equation}
In this expression, $y^2$ is a continuous function of $l$, and thus of $u$.
Furthermore, $y^2$ is a smooth, i.e.\ arbitrarily often differentiable,
function of the initial values $K_0$ and $y_0$ for any
finite $l$, see Fig.~\ref{fig:tra}.
Thus $y^2$ is a smooth function of temperature for finite $l$ ($u>0$).
On the other hand, $y^2$ is {\it not\/} a smooth, or even continuous,
function of $T$ for $l=\infty$ ($u=0$) since it has a jump
at $T_c$. The integrand in Eq.~(\ref{n.def3}), however, is
smooth in $T$ even for $u=0$.
Thus we have an integral over a finite interval over a function that is
continuous in the integration variable and smooth in the
external parameter $T$. Under these suppositions
the integral $nr_0^2$ is a smooth function of temperature even
at the transition. This behaviour is connected with the observation
that the phase transition is of infinite order \cite{Halp}.

This result is in agreement with BKT theory, where, however, one has
to introduce a cut-off $\xi_+$ to obtain it. Since the definition of $\xi_+$
is somewhat arbitrary, as discussed in Sec.~\ref{sec:intro},
the argument given here appears to be based on firmer grounds.

To obtain numerical results for the total density $n$, the generalized
Kosterlitz equations (\ref{Ko1a}) and (\ref{Ko2}) are integrated numerically
up to $l=20$, corresponding to a system size of $e^{20} r_0$,
for various values of $T/q^2$ and $E_c/q^2$. We assume $N_0=1$.
The results
for $y$ are used to approximate the integral in Eq.~(\ref{n.def2}).
The results are plotted in Fig.~\ref{fig:dens}. As expected, the
density increases with temperature and decreases with
increasing core energy. Note that the approach presented here is not
limited to $T/q^2<1/4$, where $1/4$ is the maximum value of $T_c/q^2$.
The density starts to increase below $T_c$, but the larger part of the
increase takes place above $T_c$. As expected, there is no feature at
$T_c$ (open circles in Fig.~\ref{fig:dens}). For high temperatures the density
saturates. This saturation and in fact the convergence of the density for
$T>T_c$ are due to the geometric correction of Sec.~\ref{sec:geo}.
The temperature dependence of the total density is, in principle,
experimentally accessible through, e.g., flux noise \cite{Rog} and
NQR \cite{App} measurements on superconducting films.

One may ask whether the theory is still applicable for $T>T_c$, where the
average pair size approaches the typical distance between neighboring
pairs. If both quantities are of the same order the decomposition of
the vortex system into small pairs becomes meaningless.
To address this issue we compare
the pair separation $r_p$ with the average pair size $\langle r\rangle$.
The pair separation is just $r_p=1/\sqrt{n}$. The pair size is
\begin{equation}
\langle r\rangle = \frac{1}{n} \int_{r_0}^\infty\!dr\,r\,\nu(r)
  = r_0\frac{\int_{r_0}^\infty dl\,e^{-l} y^2(l)}
    {\int_{r_0}^\infty dl\,e^{-2l} y^2(l)} .
\end{equation}
This quantity can be calculated similarly to the pair density. The
result for $E_c/q^2=0.5$ is depicted in
Fig.~\ref{fig:scal}. The important point is that $\langle r\rangle$
(dashed line) remains small compared to $r_p$ (solid line)
even in a significant temperature range
above $T_c$. In this regime large pairs are ``broken'' in the sense that the
interaction takes on a screened form for large distances, but nevertheless
the pairs do not overlap considerably and the decomposition of the vortex
system into pairs is justified. Therefore,
the theory is applicable
in the proximity of $T_c$, in addition to the low-temperature phase.
The inset of Fig.~\ref{fig:scal} shows the corresponding
quantities for $E_c/q^2=0.05$. One sees that the range of applicability
for $T>T_c$ is narrower in this case, but it still exists.

We thus have reason to believe that the extended theory is valid in a
temperature interval around $T_c$ for arbitrary core energy $E_c$, and,
therefore, that the transition is of infinite order for any $E_c$. This
result contradicts Refs.~\cite{MiWa,Zhang,Ding}, which predict a
first-order transition for small $E_c/q^2$. However,
we cannot exclude the possibility that terms of even higher order
in the Kosterlitz equations bring about such a transition or that the
expansion for small $y$ does not work at all for small $E_c/q^2$.

To conclude, straightforward physical models have been presented
for higher order terms in the Kosterlitz equations.
These terms are of a form similar, but not identical, to
earlier predictions \cite{NK,Halp}. A geometric correction,
which ensures the correct enumeration of pairs, leads to the
disappearance of the unphysical divergence of the total pair density for
$T>T_c$. Starting from an exact expression for the dielectric constant,
the polarization of the vortex system and local field effects are
taken into account, leading to a stabilization of the
high-temperature phase.
The behaviour of the stiffness constant and the correlation length
close to $T_c$ is
identical to the predictions of BKT theory so that experiments
on these quantities, e.g.\ measurements of the nonlinear resistance of
superconducting films, cannot distinguish between the two approaches.
The employment of the Clausius-Mossotti formula in Sec.~\ref{sec:pol} does not
lead to easily testable predictions. However, it is comforting to know that
the next higher order term in the equation for $K$ does not destroy the BKT
behaviour. The enumeration rule of Sec.~\ref{sec:geo}, on the other hand, leads
to new quantitative predictions for the total pair density even above $T_c$,
which can, in principle, be tested by flux noise and NQR techniques.
In short, it has been shown that the BKT approach can be extended in such a
way that the temperature regimes below and above $T_c$ are treated on equal
footing.

\section*{Acknowledgements}

The author wishes to thank J. Appel and T. Wolenski for interesting
discussions. Financial support by the Deutsche Forschungsgemeinschaft is
acknowledged.

\newpage
\section*{Figures}

\begin{figure}[h]
\caption{Trajectories of the generalized Kosterlitz
equations (\protect\ref{Ko1a}) and (\protect\ref{Ko2}) for $E_c/q^2=0.5$ and
varying $T/q^2$ (thin solid lines), and for $E_c/q^2=0.2$ and
varying $T/q^2$ (dashed lines).
The separatrix is denoted by the thick solid line, the bound-pair phase
is below that line. The dotted lines denote initial values ($l=0$)
for $E_c/q^2 = 0.2$ and $0.5$, respectively, and varying temperature.}
\label{fig:tra}
\end{figure}

\begin{figure}[h]
\caption{The total pair density $n$ as a function of $T/q^2$ for
$E_c/q^2 = 0.02$, $0.05$, $0.1$, $0.2$, $0.5$, $1$ (from left to right).
The open circles denote
the respective values of $T_c/q^2$ and the corresponding densities.}
\label{fig:dens}
\end{figure}

\begin{figure}[h]
\caption{The average pair size $\langle r\rangle$ (dashed line) and
the average separation between neighboring pairs (solid line) for
$E_c/q^2=0.5$. The filled triangle denotes $T_c$. The inset shows the same
quantities for $E_c/q^2=0.05$.}
\label{fig:scal}
\end{figure}

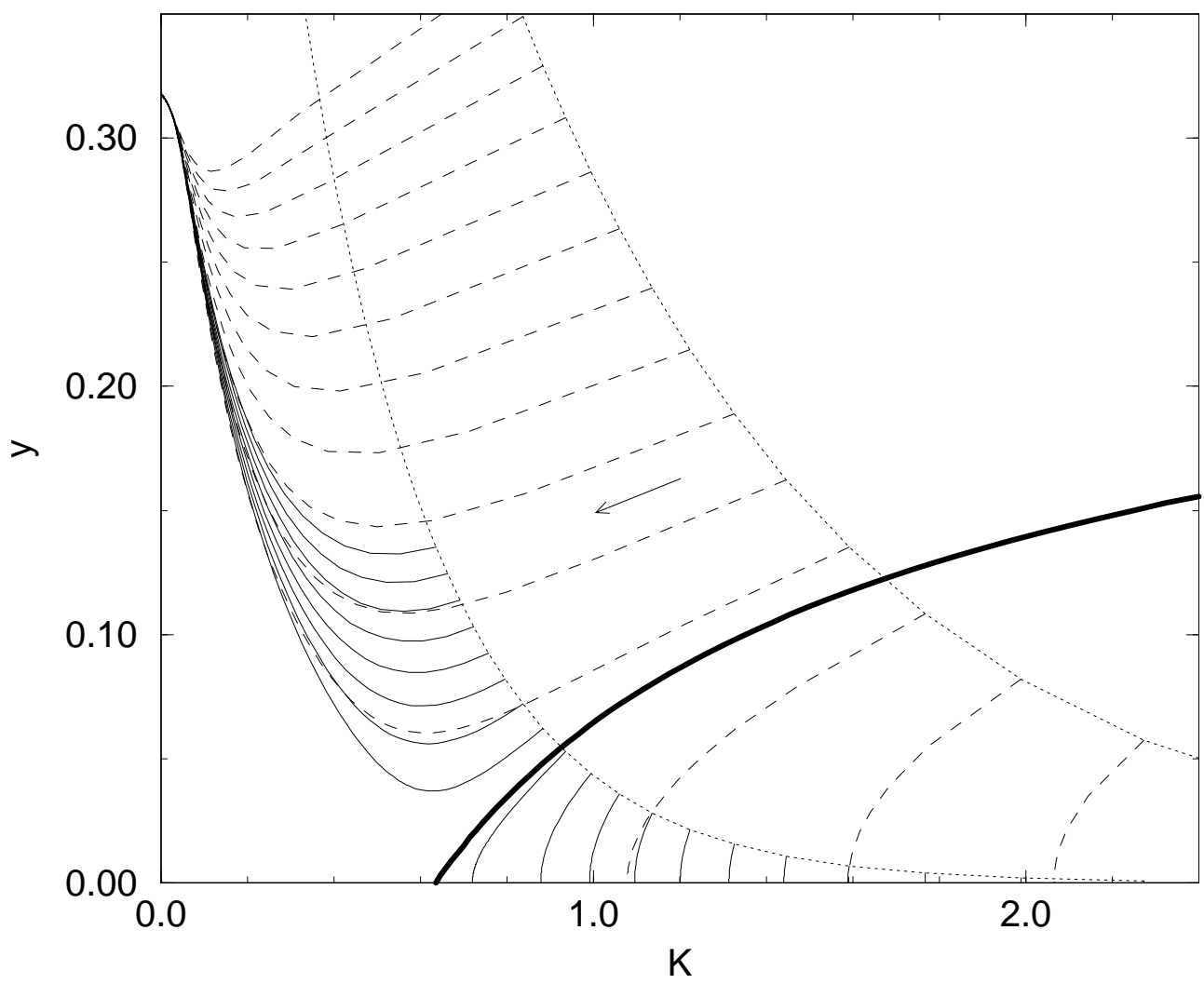
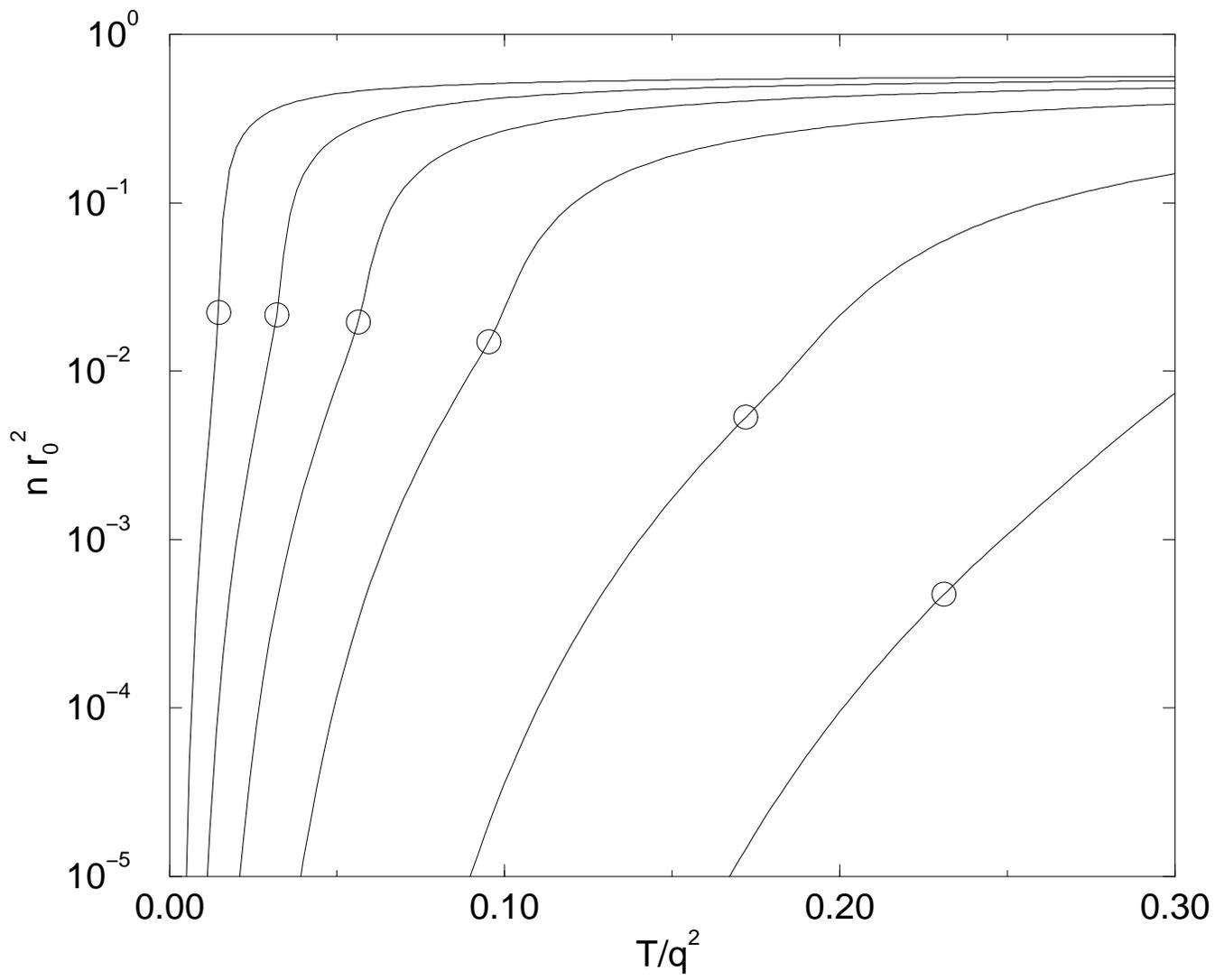
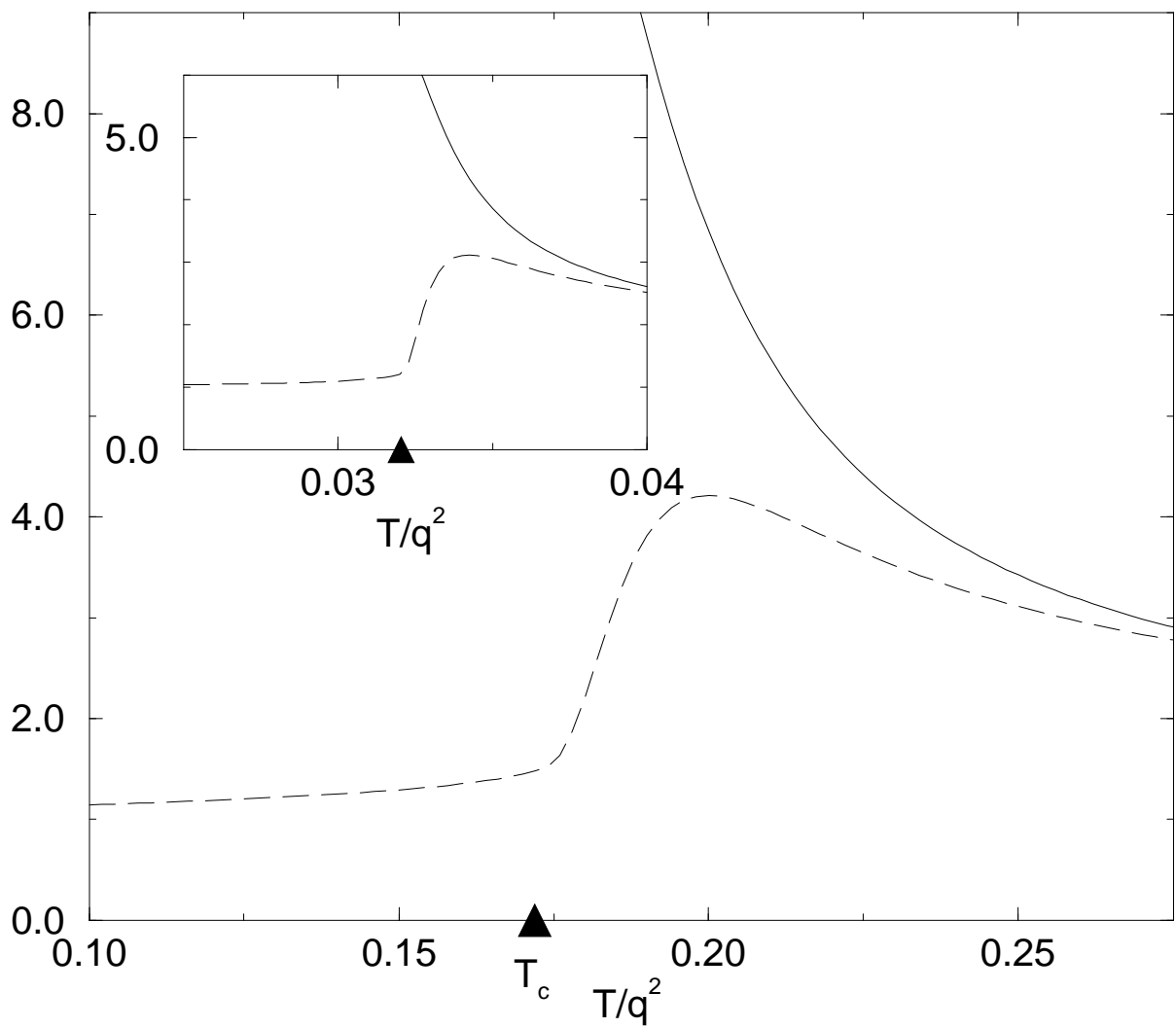
\end{document}